\documentclass[sigconf, 10pt]{acmart}

\AtBeginDocument{%
  }

\usepackage{amsmath,amssymb,amsfonts}
\usepackage{algorithmic}
\usepackage{graphicx}
\usepackage{textcomp}
\usepackage{xcolor}
\usepackage{tikz}
\newcommand{\circled}[1]{\tikz[baseline=(char.base)]{\node[shape=circle,draw,inner sep=1pt] (char) {#1};}}

\newcommand{\sys}{SpeechShield}
\newcommand{\slurp}{SLURP-clean}
\newcommand{\slurpMix}{SLURP-Mix}
\newcommand{\cloud}{AllOffload}
\newcommand{\preechOrg}{Preech (Original)}
\newcommand{\preechWhisper}{Preech (Whisper)}
\newcommand{\WhisperBase}{OnDevice-Whisper (Base)}
\newcommand{\WhisperTiny}{OnDevice-Whisper (Tiny)}
\newcommand{\ourszeroshot}{Ours-zeroshot}
\newcommand{\oursfinetuned}{Ours-finetuned}

\copyrightyear{2025}
\acmYear{2025}
\setcopyright{cc}
\setcctype{by}
\acmConference[SEC '25]{The Tenth ACM/IEEE Symposium on Edge Computing}{December 3--6, 2025}{Arlington, VA, USA}
\acmBooktitle{The Tenth ACM/IEEE Symposium on Edge Computing (SEC '25), December 3--6, 2025, Arlington, VA, USA}
\acmDOI{10.1145/3769102.3770609}
\acmISBN{979-8-4007-2238-7/2025/12}

\begin{document}

\title{Safeguarding Privacy in Edge Speech Understanding with Tiny Foundation Models}



\author{Afsara Benazir}
\affiliation{%
\institution{Department of Computer Science}
  \institution{University of Virginia}
  \city{Charlottesville}
  \state{VA}
  \country{USA}
}
\email{hys4qm@virginia.edu}

\author{Felix Xiaozhu Lin}
\affiliation{%
\institution{Department of Computer Science}
  \institution{University of Virginia}
  \city{Charlottesville}
  \state{VA}
  \country{USA}
}
\email{felixlin@virginia.edu}


\begin{abstract}
Robust speech recognition systems rely on cloud service providers for inference. It needs to ensure
that an untrustworthy provider cannot deduce the sensitive content in speech. Sanitization can be done on speech content keeping in mind that it has to avoid compromising transcription accuracy. Realizing the under-utilized capabilities of tiny speech foundation models (FMs), for the first time, we propose a novel use: enhancing speech privacy on resource-constrained devices. 
We introduce \sys{}, an edge/cloud  privacy preserving speech inference engine that can filter sensitive entities without compromising transcript accuracy. We utilize a timestamp based on-device masking approach that utilizes a token to entity prediction model to filter sensitive entities. Our choice of mask strategically conceals parts of the input and hides sensitive data. The masked input is sent to a trusted cloud service or to a local hub to generate the masked output. The effectiveness of \sys{} hinges on how well the entity time segments are masked.
Our recovery is a confidence score based approach that chooses the best prediction between the cloud and the on-device model.

We implement \sys{} on a  64 bit Raspberry Pi 4B. Experiments show that our solution leads to robust speech recognition without forsaking privacy. \sys{} with $< 100$ MB memory, achieves state-of-the-art (SOTA) speech transcription performance while filtering about 83\% of private entities directly on-device. \sys{} is 16x smaller in memory, 3.3x faster and 17x more compute efficient than prior privacy preserving speech frameworks and has a relative reduction in word error rate (WER) by 38.8-77.5\% when compared to existing offline transcription services.

\end{abstract}


  

\maketitle
\section{Introduction}
\label{sec:intro}
Speech is a pervasive interface to embedded devices.
A key speech task is automatic speech recognition (ASR), 
transcribing voices to textual sentences, 
which further feed into various AI tasks such as instruction-taking LLMs, robot manipulations, and IoT actions. 
Transcribing natural, non-trivial utterances (e.g. "Pick up the red cube on the table and put in the basket") necessitates deep ML models each having
GBs of memory and tens of GFLOPs per input second \cite{arora2022two,radford2022robust}. 
Facing the constraints, many embedded devices choose to 
offload all voices to the cloud \cite{microsoft_azure_speech}. 

\begin{figure}
\centering
\includegraphics[width=\columnwidth]{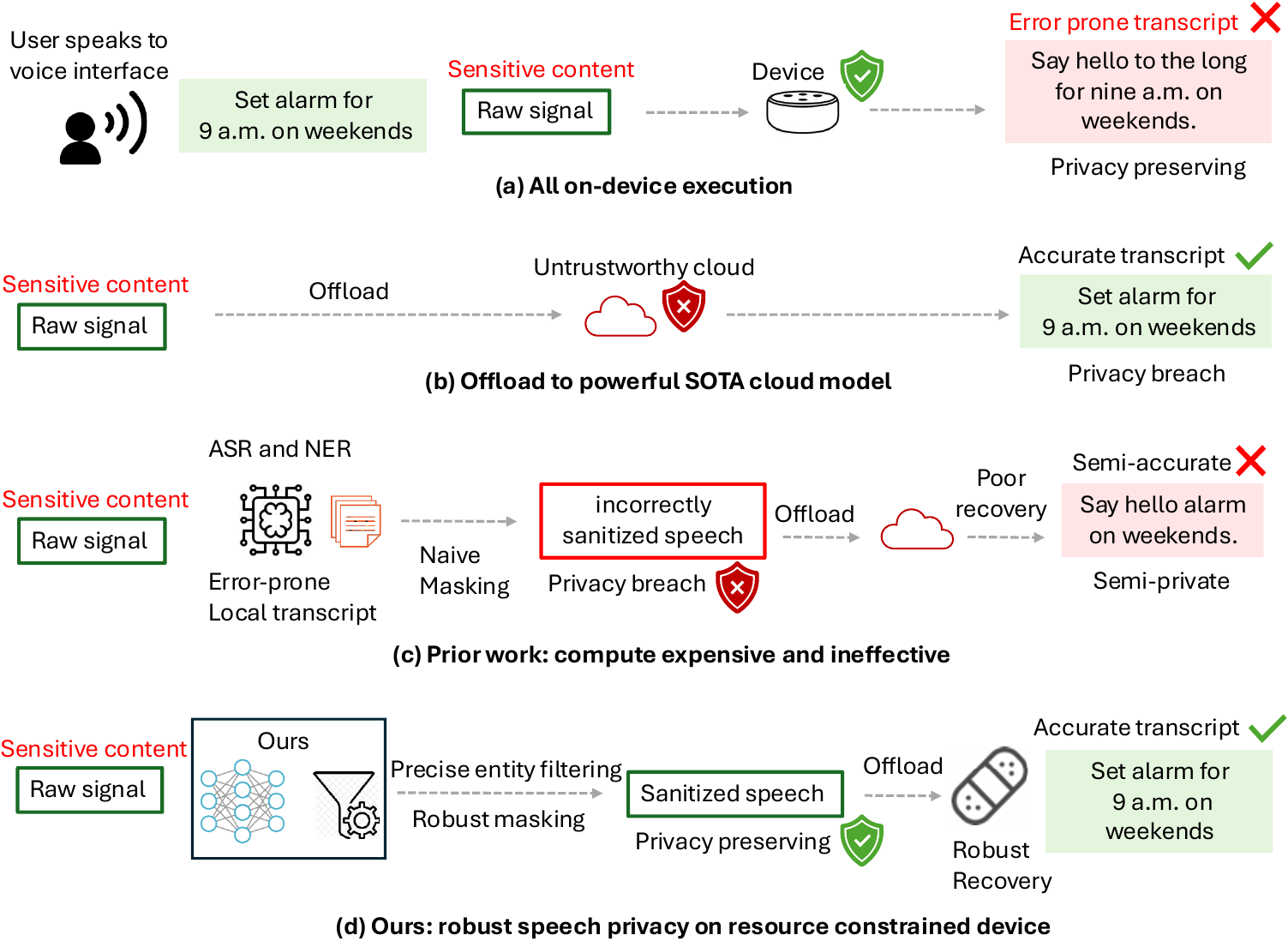}
    
    \caption{Untrustworthy cloud and risk of privacy. (a) All on-device execution fails at ASR inference while (b) leaks all sensitive information. (c) shows prior privacy preserving speech transcription framework (d) Ours is a secure timestamp based speech filter on mobile devices.}

    \label{fig:pr}
\end{figure}

This paper focuses on mitigating privacy leak from offloaded voice inputs to cloud ASR services (ref. Figure \ref{fig:pr}). 
Specifically, we enable a resource-constrained device to mask \textit{named entities} in a voice input, before offloading the input to the cloud. 
Named entities, such as names of people, location, dates and time etc are regarded as the major sources of privacy leak as shown in extensive research~\cite{iqbal2023tracking,saade2019spoken, brown2022does}. 
Compared to obfuscating a speaker's identity (\textit{paralinguistic} knowledge such as gender, emotion or race) or voiceprint privacy
\cite{dias2018exploring, aloufi2019emotionless, granqvist2020improving, gong2017crafting, stoidis2021protecting, jaiswal2020privacy, wu2021understanding, kroger2020privacy},
precise masking of named entities in an utterance is much more difficult 
-- the device would need \textit{semantic} knowledge of the utterance that typically requires a deep ML model. 
This is why prior privacy-preserving systems in speech privacy (ref. \S\ref{sec:related-work}) either have to run a model as large as 1.5 GB on device defeating the benefit of cloud offloading~\cite{ahmed2020preech}, 
or dodge the problem altogether by making the cloud infer only speech \textit{intents} instead of \textit{transcripts}~\cite{cai2024lightweight}, which prevents downstream tasks such as text/image generation. 
Hardware based keyword filtering such as \cite{olade2019smart} introduces a power heavy, computationally expensive intermediary device between the user and smart speaker; \cite{brasser2018voiceguard} runs inference in a memory intensive trusted execution environment (TEE) - our overhead on commodity hardware is significantly lower (Detailed in  \S\ref{sec:o}).


\textbf{Our insight: a tiny foundation model as privacy guard}
The emerging foundation models (FM) for speech, 
such as Whisper \cite{radford2022robust} are shown to generalize well to various voice inputs. 
Even the tiny variant of the model family, e.g. Whisper-Tiny (75 MB), 
was pretrained on 680,000 hours of voice data. 
These tiny variants, however, were largely ignored in prior work: 
due to their low model capacity, the accuracy is simply too low to be usable (see Table \ref{tab:bwer}). 
We, however, identify them as the long-missing building block for speech privacy: 
thanks to their colossal training data, 
tiny FMs have adequate knowledge for labeling privacy-sensitive entities. 

To realize the idea, we address the following challenges.
\textbf{What we understand by sensitive content?} We consider the widely accepted 18 categories of named entities (person, organization, date, quantity, product etc.) from OntoNotes \cite{weischedel2011ontonotes} as sensitive.  Context in an utterance such as - '\textit{meeting}' in '\textit{I have a meeting \underline{tomorrow}}' or \textit{'book'} in \textit{'Please book a flight to \underline{New York}'} is essential to comprehend the underlying intent behind it (Action: \textit{set\_meeting} or \textit{book\_flight})\cite{lugosch2019speech}. Hence we preserve the general context surrounding a named entity. For domain specific scenarios e.g.
healthcare/legal application, private data is more nuanced,
encompassing intricate named entities and domain-specific
terminology (e.g. name of a disease) requiring further sophisticated filtering. For such cases the on-device model needs
domain specific finetuning, beyond the scope of this work.

\textbf{How to locate entity positions?} Prior systems locate named entity in a spoken input based on the tentative transcript~\cite{ahmed2020preech} and naive word segmentation. 
The results are brittle, as resource-constrained devices cannot afford high-quality ASR. 
To this end, we exploit Whisper's output timestamps, predicted through alignment based on Dynamic Time Warping (DTW) \cite{sakoe1978dynamic} - a technique much more precise than naive word segmentation. We introduce a lightweight, on-device \textit{named entity recognition} (NER) model that locates the entity timestamps from output tokens allowing precise masking of entities based on these timestamps.

\textbf{How to sanitize a voice input?}
Simply muting the timespan of named entities does not work: 
as the cloud ASR model expects an integral voice input, it tends to hallucinate on such an input with intermittent silence spans and abruptly stops after the first observed silent span. 
To this end we experiment with different choices of mask (detailed in \S\ref{sub:di}); among the possible choices, random white noise serves as an optimal sanitizer, likely because the cloud is trained to make predictions even from noisy segments. 

\textbf{How to recover transcripts after masking?}
Processing a sanitized voice, the cloud returns a transcript that contains erroneous text tokens, in and around timespans of named entities. 
To substitute these tokens with the right ones for named entities, the named entities in edge inference needs to be \textit{patched} with the cloud inferred transcript. This can be done by aligning the timestamps of two predictions, which is difficult to achieve. Another approach is to use the token confidence score of the two predictions to determine the most suitable token among cloud and edge, ensuring higher transcription accuracy. Methodology is detailed in \S\ref{sec:e2e-sys-des}.

\textbf{How to integrate in an edge/cloud ASR pipeline?}
Tiny FMs such as Whisper-Tiny \cite{radford2022robust} can process spoken inputs with varying level of confidence score per token. For inference with high confidence, local processing is acceptable. 
The integrated pipeline consists of an edge processing unit consisting of a tiny speech FM followed by a lightweight entity label classifier. Entity filtering, speech sanitization, patching all are done on device; cloud is invoked for inference on the filtered speech only.

\textbf{Our system.} \sys{} is 
an edge/cloud ASR system that automatically masks spoken entities before offloading to cloud, while still generating high-quality transcripts. 
\sys{} distinguishes from prior work with two notable designs (1) It hides entities in utterance combining on-device token based entity filtering with timestamp based masking (2) It recovers the complete ASR transcript from masked speech on-device using a unique confidence score based patching.

\sys{} trains a lightweight \textit{named entity recognition} classifier to filter sensitive entities on-device. It uses decoder output tokens and cross-attention-based word-level timestamps to predict entity positions without relying on erroneous edge ASR. An off-the-shelf Whisper-Tiny and Whisper-Large-v3 serve as the edge and cloud, respectively. \sys{} fundamentally differs from prior privacy preserving framework \cite{ahmed2020preech, cai2024lightweightprotectionprivacyoffloaded} because of its superior filtering and patching approach. 

Our end \textbf{goal} is (1) to prevent sensitive content i.e named entities from being exposed to the cloud and (2) improve upon the transcription accuracy compared to prior work.
The recovered transcript can serve several purposes such as transcription of real-time meetings, voice call logging, intent-based action taking in edge
interfaces (ref. \S\ref{sec:usability}).

Our training, inference and data processing scripts are publicly available at \href{https://github.com/afsara-ben/whisper-ner}{https://github.com/afsara-ben/whisper-ner} to ensure reproducibility.


\textbf{Results.}
We prototype \sys{} on an embedded device and a backend server and extensively  evaluate on challenging voice datasets. 
Running in real-time and with less than 100 MB memory, 
\sys{} is able to filter about 83\% of named entities in voice inputs, 1.6x-2.7x higher than prior systems with similar privacy goals~\cite{ahmed2020preech}. 
\sys{} requires 16x less memory, 17x less compute and is 3.3x faster; 
meanwhile, \sys{}'s transcription accuracy is on par with state of the art (SOTA) results,  seeing negligible degradation. Compared to available on-device transcription services, \sys{} demonstrates a relative
improvement in word error rate (WER) by 38.8-77.5\% on real-world evaluation datasets (\S\ref{sec:e2e}).



 

 

\textbf{Contribution.}
We make the following contributions. 

\begin{itemize}
	\item We showcase a novel use of tiny foundation models for privacy: to recognize and then filter named entities in user speech, before offloading the speech to the cloud. 
	
    \item Correspondingly, we propose novel techniques suitable to embedded devices: a lightweight classifier to label the time boundaries of entities and properly mask them; confidence score based recovery to patch in edge generated entities with cloud-generated transcripts.   
    
    \item We further apply an array of optimizations, including named entity masking strategy, model tuning, edge model threshold etc. which boost the system accuracy and efficiency significantly. 
\end{itemize}

The remainder of this paper is organized as follows. Section \ref{sec:motiv} discusses the background and motivation for \sys{}. Section \ref{sec:overview} \& Section \ref{sec:design},\ref{sec:recovery} describes the overview and design space. 
Section \ref{sec:opt} and Section \ref{sec:eval} present a varied number of optimizations, experimental setup and evaluation results.
\section{Background and Motivation}
\label{sec:motiv}

\subsection{Background}
\label{subsec:back}
\paragraph{Spoken language understanding} (SLU) has been widely researched in recent days. From the conventional cascading automatic speech recognition (ASR) and natural language understanding (NLU) pipeline which is resource intensive, error-prone and slow, recent models follow an end to end pipeline \cite{qin2021survey} where inference is performed directly from speech embeddings to text transcription.


\paragraph{Named Entity Recognition} (NER) involves identifying and extracting entities from text, such as names, locations, dates etc. \cite{etzioni2005unsupervised}.
It involves text preprocessing - text tokenization and normalization; using a pretrained NER model to perform feature extraction, that can be lexical, syntactic or contextual features. Standard NER performance relies on the robustness of text transcripts.




\paragraph{Whisper}
The Whisper \cite{radford2022robust} model family from openAI is a robust speech recognition architecture consisting of different model sizes and trained on 680,000 hours of speech data. 
The Large-v3 model ($>1.5$B parameters) has an average WER of 0.13 (0.03–0.25), while Whisper-Tiny (37M parameters) has an average WER of 0.24 (0.07–0.58) across 14 speech datasets. Performance of different whisper models on our evaluation set is in Table \ref{tab:bwer}.

Obtaining the performance of Large-v3 at the cost of running Whisper-Tiny can be alluring but seems impossible. As the model size decreases, its WER also increases - this trend aligns with the scaling law of large models \cite{kaplan2020scaling}. 

\paragraph{Token-level timestamps}
Whisper extracts token-level timestamps using the cross-attention pattern and alignment based on Dynamic Time Warping (DTW) \cite{sakoe1978dynamic} in order to extract word boundary information. DTW finds the optimal alignment indices for the text tokens based on the attention weights. The cross attention weights are different for different whisper models on same input, thus predicted timestamps may differ by 100-400ms for the same generated tokens by local and cloud model.
The start and end times for each token are calculated based on the token boundaries and the time indices are obtained from DTW. Additionally, the average probability (confidence score) of text tokens within each word is calculated.

\subsection{Motivation}

We observe that tiny FMs are notoriously bad in general ASR tasks, but it is adequate enough to tag entities from the generated tokens. 
For example  – tiny FM predicts the utterance "\textit{His wife's birthday party is at 10 PM.}" as "\textit{Is one for the apartment at 10 PM.}" 
Although the ASR transcript is inaccurate, the entity time boundary is consistent. Auxiliary outputs in an inference such as token level timestamps and token level confidence score can be utilized for segmenting entity boundaries and transcript retrieval respectively. We recognize that entity words can only be transcribed by the edge model, that might be error-prone sometimes.

\subsection{Application}
\label{sec:usability}
Voice user interfaces (VUI) are integrated into almost everything: from hand held gadgets such as smart speakers, tablet, earphones to appliances such as TVs and set-top boxes to cars \cite{cheng2022personal}. Cloud can misuse user’s voice search history, voice command history etc.
We allow these voice assistants to be functional without compromising sensitive information.
Our effectiveness extends beyond entity recognition to the entire transcript. 

A plausible use case scenario of our framework is in transcribing accurate meeting notes, personal voice call logging, using the transcript for intent based action taking in edge speech interfaces etc. -- all on-device, ensuring privacy. 

We target automatic speech recognition (ASR) on mobile devices that can range from smartphone devices (with 4GB CPU ram) to any user interface devices such as voice assistants, smartwatch, home automation devices (with less than 100 MB RAM). Our test platform is detailed in \S\ref{sub:m}.

\subsection{Our System Model}



We assume that the device has an online part - a cloud service or a nearby hub, which we refer to as “cloud” for short. The user (victim) offloads spoken inputs to the cloud (adversary) for inference. The cloud hosts a state-of-the-art (SOTA) ASR model. 

Our system model consists of an offline and online phase. The client speaks to the device which only offloads the filtered utterance containing no named entities. The on-device model transcribes the entity segments and saves them locally. Once online, the cloud retrieves the transcript from the masked speech, sends back to device and with a clever recovery technique, the two transcripts are merged to form a complete sentence.  




\subsection{Threat Model}

Our threat model agrees with prior work \cite{wang2024privacy, aloufi2020privacy} where the users provide commands to the edge device that revokes the cloud service providers to maximize
ASR performance. While using these speech-based services, users grant complete access to their recordings. The expectation is that sensitive information in the data should be protected and that the data will only be used for the intended task (e.g. execution of a voice command). Use of the data for further analyses is an act of violation, as it allows the service providers to generate targeted content. 

\subsection{Side-channel attacks}
Following prior work on speech privacy \cite{backstrom2023privacy}, our focus is primarily on information isolation of predetermined categories within our edge resource budget. 
We acknowledge side-channel threats under a semi-honest cloud (e.g. honest-but-curious attacks \cite{bonawitz2017practical}), but these protections are orthogonal to our contribution, and practical mitigations are limited.

For example, offloading short ( $\leq$ 6s) speech segments across multiple providers introduces synchronization errors, while injecting randomized speech into non-entity segments (effectively treating the non-entity segments as sensitive) was rejected owing to (1) Generating randomized prompt tokens on the client side requires a text generator and text-to-speech (TTS) model that has massive memory/compute overhead—the smallest GPT-2 text generator alone has 124M parameters—and (2) injecting a high amount of out-of-order words will affect the cloud model's predictions, as it is trained to process coherent sequences of tokens.

Adversarial attacks e.g. spoofing are beyond the scope of this work.
Although not specifically delineated - our pipeline inherently serves as a pseudo ‘Data Reconstruction Attack’, where the cloud is given complete access to the masked data and it tries to reconstruct the complete ground truth from the masked speech.

\subsection{Privacy Guarantee}
A theoretical privacy guarantee such as differential privacy (DP) \cite{dwork2006differential} that adds probabilistic noise to hide individual data is non-applicable in our scenario as the sensitive segments are designed to not be offloaded. Compared to DP, our trust model is more \textit{deterministic} – masking replaces the sensitive data with an alternate (dummy speech segments or plain white noise). DP would be essential in the scenario where the sensitive speech segments are shared with the cloud (for entity specific downstream task). Adding differential privacy or similar privacy guarantee also comes with an implementation burden. In the baseline Preech \cite{ahmed2020preech} - an on-device GPT-2 model is used to ensure DP resulting in significant memory and compute overhead.

We provide a deterministic guarantee of \textit{entity indistinguishability} analogous to \textit{information-isolation} in \cite{backstrom2023privacy}.
Given an utterance $x = (E, N)$, where $E$ denotes the sensitive entity spans and $N$ the non-entity spans, the cloud observes only a sanitized view $V(x) = f(N)$ after masking. For any two utterances $(E_1, N)$ and $(E_2, N)$ that differ only in their entity spans, the observable views remain identical, \[V(E_1, N) = V(E_2, N)\] i.e the cloud cannot distinguish which entities were present, as both yield the same observable data.




\section{\sys{} Overview}
\label{sec:overview}


\sys{} is a pipelined framework consisting of \circled{1} an on-device tiny FM that generates tokens along with their timestamps \circled{2} an on-device lightweight entity prediction classifier that assigns each token a label (entity or non-entity) \circled{3} a mask configurator that precisely masks the entity segments based on their time boundaries \circled{4} a robust recovery protocol to patch in edge and cloud inferred segments.

\begin{figure*}[t]
\centering
\includegraphics[width=1.9\columnwidth]{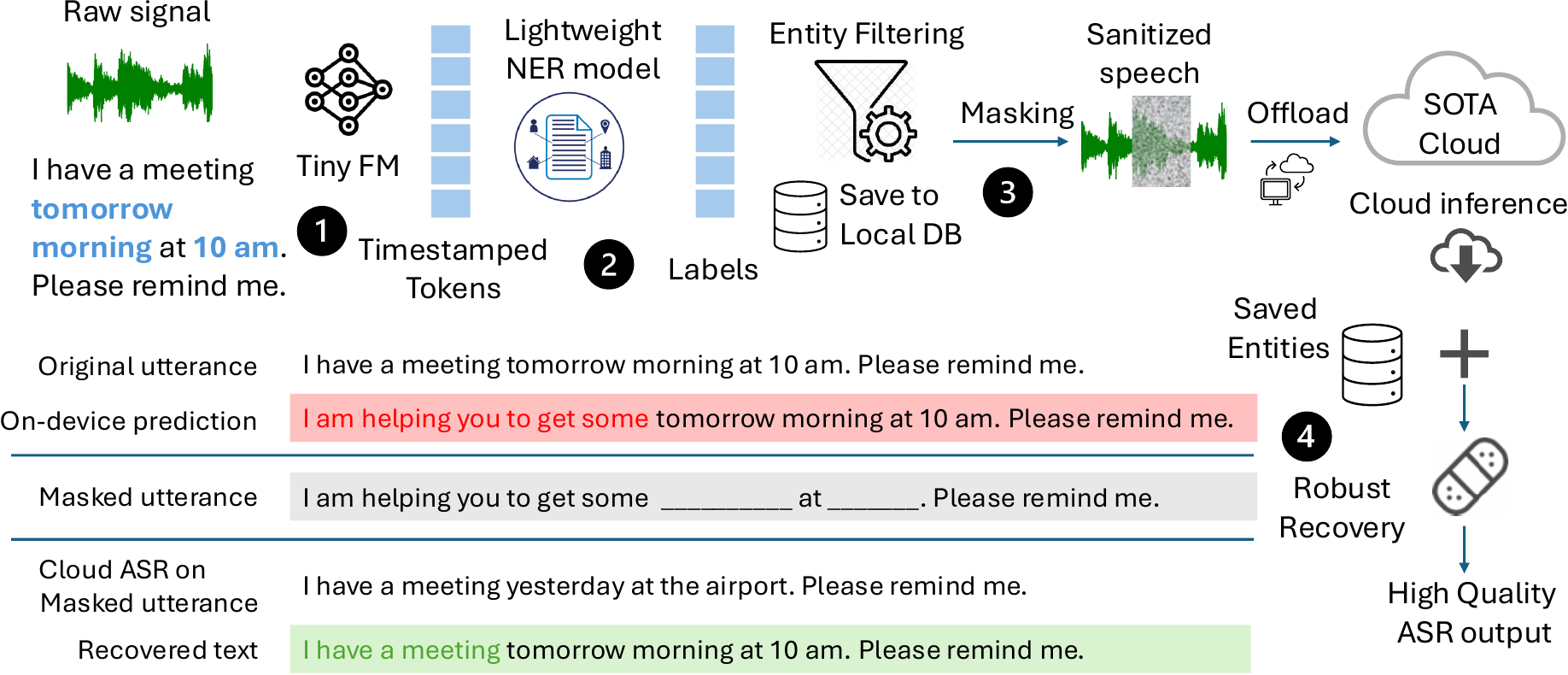}
    
    \caption{Overview of \sys{}. Green text represents successfully recovered words. Bold blue text are sensitive, private information. Red text is erroneous prediction.
    }

    \label{fig:o}
\end{figure*}

\textbf{Operations.}
The original raw speech is processed by the on-device tiny FM that generates discrete tokens along with their time boundaries (timestamps). The on-device entity prediction classifier then assigns a label (either 0=non-entity or 1=entity) to each output token, so each timestamped segment has a corresponding label. Although the tentative on-device transcript might be incorrect, our classifier can handle a binary classification task from limited token information. The text transcript of segments with label "1" is saved locally to be used in recovery phase later on. 

The segments with label "1" are "masked" with a choice of available masks. The masked speech is offloaded to the cloud for inference, where a SOTA cloud ASR model performs inference on the masked input. The edge device receives this incomplete transcript and tries to recover a complete transcript. It patches in the entity segments either using a timestamp based alignment approach or confidence score based approach (\S\ref{sec:e2e-sys-des}). The workflow is illustrated in Figure \ref{fig:o}.


\subsection{Design Implications}
\label{sub:di}
Our choice of design separates us from existent work in speech privacy. 

\paragraph{\textbf{Entity prediction classifier}}
We want our on-device classifier to be lightweight yet reliable in labeling each token to a binary class (entity or non-entity). The added complexity of a conventional named entity recognition model is unnecessary for our purpose. A traditional recurrent neural network (RNN) \cite{sherstinsky2020fundamentals} can handle this binary classification effectively; in turn a transformer \cite{vaswani2017attention} based adapter or joint named entity and speech transcription model \cite{ayache2024whisperner} is too expensive for local inference and would be an overkill.


\paragraph{\textbf{Masking configuration}}
\begin{figure}[t]
\centering
\includegraphics[width=\columnwidth]{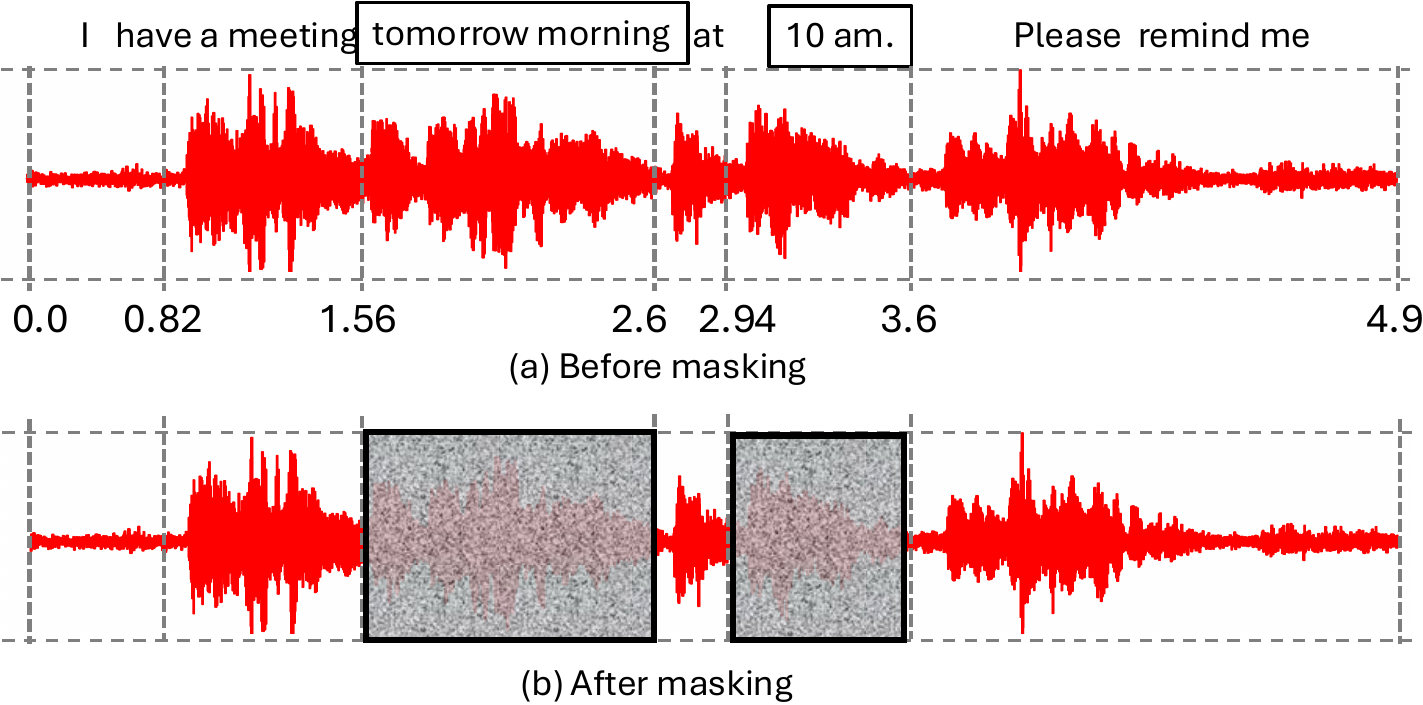}
    
    \caption{Masking configuration}

    \label{fig:mc}
\end{figure}

It is crucial to decide which segments to mask and how to mask. Following prior work \cite{ahmed2020preech, cai2024lightweight, wang2024privacy} we conceal named entities in spoken input.

\textit{Which part to mask?} Prior work identified entities from on-device error prone text transcripts and segmented word 
boundaries using silence as a separator. This approach can be frail.
Instead we predict named entities using our token to entity prediction classifier and mask the content of the respective token's timestamp. Our masking strategy completely relies on timestamps and is more accurate. 

\textit{How to mask?}
Incorrect or insufficient masking configurations can leak sensitive information and compromise privacy. We partition the spoken input into discrete segments based on their token timestamps and only replace the sensitive segments with a choice of \textit{mask}; rest of the audio is kept intact (ref. Figure \ref{fig:mc}). 

\textit{Choice of mask?} Incorrect or insufficient masking configurations can leak sensitive information and compromise privacy. We partition the spoken input into discrete segments based on their token timestamps and only replace the sensitive segments with a choice of \textit{mask}; rest of the audio is kept intact (ref. Figure \ref{fig:mc}). We explore different choice of masks: silence, white noise, melody \& dummy word injection (\S\ref{sub:m}).

Empirically, we find that white noise is the better mask configurator; cloud language models can recover well from noise-masked inputs.

\paragraph{\textbf{Choice of Transcription Recovery}}
The cloud returns pseudo-transcripts on the masked segments that needs to be replaced by the locally saved entity segments. This patching (or recovery) approach can be (1) timestamp based that works best when the edge and cloud prediction are aligned but dissimilar and (2) confidence score based, that chooses the most credible tokens. Elaborated in 
Section \ref{sec:recovery}.

\section{Speech filtering and masking}
\label{sec:design}
\label{sec:e2e-sys-des}
\subsection{Problem Statement}
Similar to \cite{wu2021understanding}, we define client-side privacy: 
For a user $u$ and utterance $X$ having private entities $Y$, we want to learn an encrypted signal $X^\prime$ from $X$ under the privacy-preserving function $F(\theta) :X \longrightarrow X^\prime$ with parameters $\theta$. $F(\theta)$ should learn an encrypted/obscured signal $X^\prime$ ensuring efficiency, utility and privacy. One should not be able to decode $Y$ from $X^\prime$. $$\forall y \in Y \mbox{ , } F(X)=X^\prime : X^\prime \cap Y = \emptyset$$


\subsection{Design}
\textbf{(1) Timestamped token generation.}
The on-device tiny FM decoder auto-regressively predicts tokens, conditional on previously predicted tokens and the encoder hidden states. Word-level timestamps obtained through cross attention weight and DTW alignment is more accurate and compute efficient than using a phoneme level aligner as in \cite{bain2023whisperx}. Our tiny FM generates a set of timestamped segments $E$ where each segment $e \in E$ is defined by a start time $e_{start}$, end time $e_{end}$ and a numerical token value. Each segment can consist of one or more tokens.

\textbf{(2) Entity prediction.}
The entity prediction classifier maps the sequence of decoder output tokens and assigns them a label (entity or non-entity). 
We define the entity segments $T = t_{1} ... t_{n}$;  $t_{x}$ is a tuple with start and end points $[t_{start}, t_{end}]$ and its associated entity $t_{ent}$; $t_x=\{t_{start}, t_{end}, t_{ent}\}$.

\textbf{(3) Timestamp based content masking.}
We iterate over $E$ in order to filter out the entity segments, $T$. We eliminate using a preferred choice of mask.
If $S(t)$ represents the input speech signal and  $M(t)$ is a function defining the mask:
\begin{equation}
    S(t)^{\prime}=S(t)\times M(t)
\end{equation}
$S(t)^{\prime}$ is the masked speech we offload to cloud that hosts a robust ASR model. We keep a local copy of the entity segments, $T$ - to be used later.

discuss the design complexity, explain why a separate classifier is necessary for NER, and justify why the end-to-end approach was discarded.

\subsection{Implementation}
\paragraph{\textbf{Architecture}}
Our on-device execution is a pipelined architecture with a Whisper-Tiny model (39M parameters in size) cascaded with our pretrained, small footprint entity recognition classifier (or NER model) having only 6.9M parameters (Ref. Figure \ref{fig:o}). 
Whisper-Tiny is a vanilla encoder-decoder transformer with 4 encoder-decoder layers, 6 attention heads and a width dimension of 384.

The on-device tiny FM can generate timestamps based on cross-attention weights (ref. \S\ref{subsec:back}), but simple model finetuning cannot perform prediction-based downstream tasks such as entity recognition; they require separate classifiers or a classifier head to map encoder representations to discrete labels \cite{baevski2020wav2vec,radford2022robust}.

We contemplated appending a classifier head after Whisper-Tiny's last decoder state to output binary labels end-to-end but discarded it as (1) using hidden states offer no latency benefit owing to autoregressive generation, (2) no ground truth exists to label hidden states directly without tokenization, and (3) a standalone classifier keeps the design modular, flexible, and fault-tolerant without compromising accuracy or latency.

Thus our classifier is a 3 layered BiLSTM architecture with a FFN and a hidden dimension of 128, the input to which is the output tokens from the last decoding stage of the whisper model. The outputs are the timestamps (start and end) of the entity tokens. 



\paragraph{\textbf{Training}}
Only the NER model is trainable; the on-device Whisper-Tiny parameters are kept frozen.
We train our lightweight on-device NER model for a binary classification task: to predict whether each decoded token in a sequence is an entity or not. 
The outputs are the timestamps
(start and end) of the entity tokens.

It has three main components: \begin{itemize}
    \item An embedding layer that converts input tokens from the last decoding stage of the on-device tiny FM into dense vector representations, having a hidden size of 128 and input size of 51865 (equivalent to the vocabulary size of OpenAI-Whisper). 
    \item 3 bidirectional LSTM (BiLSTM) layers that process these embeddings in both forward and backward directions to capture contextual information.
    \item A fully connected output layer that returns a probability distribution  for each token in the sequence. We take the \textit{argmax} of the distribution to get the highest probability. We use a 0.5 threshold; probabilities above it indicate an entity.
\end{itemize}

\textit{NER(ent.-only)} trained on entity-only utterances, but evaluated on all, highlights the need for diverse training data (\textit{NER(Ent.+non-Ent.)}) for optimal performance (\S\ref{tab:ner}).

\label{app:mt}
We train this model with hyperparameters: learning rate 0.001, 10 epochs, batch size=4. We use \textit{BCEWithLogitsLoss}
as the loss function, suitable for binary classification tasks.
Training is optimized using the AdamW optimizer, which
includes weight decay for regularization.

\section{Transcription recovery}
\label{sec:recovery}
\subsection{Problem Statement}
The cloud model runs inference on the offloaded masked input $S(t)^{\prime}$ and outputs a transcript $C$. For input segments where entities are removed or obfuscated the cloud generates a pseudo transcript; for non-entity segments, the transcription is accurate as the cloud is robust. Each segment in $C$ corresponds either to an entity $C_e$ or non-entity $C_i$. The cloud ASR output $C$ is sent back to the edge device. The goal is to form a complete transcript $\hat{T}$ by merging edge inference $E$ and cloud inference $C$, such that the transcriptions of entity segments are from $E$ and non-entity segments are from $C_i$. $C_e$ segments are discarded. This recovery task can follow two designs: either aligning the inference timestamps or choosing the token of higher confidence as discussed next.




\subsection{\textbf{Approach A: Timestamp Based Recovery}}
This is more complex than timestamp based content masking as we need to align the cloud and edge segments and \textit{'patch in'} replacements. After offload, the cloud sends back its prediction $C$, a series of text token segments $(c_1, c_2, ... , c_n)$. $T$ is the entity segments list saved locally. 

\textbf{(1) Segment Removal.}
We remove segments in $C$ based on the specified start and end times in $T$.
We iterate over $C$ and for each segment $c$, we  check if it should be removed.
A segment $c$ is removed when there is an overlap between $(c_{start}, c_{end})$ with any entry in $T$. We filter out the segments that satisfy at least one of the following conditions:
\begin{equation}
    t_{start} \leq c_{start} \leq t_{end} \mbox{ or }
    t_{start} \leq c_{end} \leq t_{end}
\end{equation}
\begin{equation}
    c_{start} \geq t_{start} \mbox{ and } c_{end} \leq t_{end}
\end{equation}
\begin{equation}
    t_{start} \geq c_{start} \mbox{ and } t_{end} \leq c_{end} 
\end{equation}
Moreover, the timestamps generated for consecutive tokens \textit{n} and \textit{n+1} have the same end and start time. This can cause excessive removal of tokens. To avoid this we right shift each start time by 10 ms and left shift end time by 10 ms.


\textbf{(2) Entity Insertion.}
A replacement $r \in R$ is characterized by its start time $s_r$, end time $e_r$ and associated token(s) $t_r$; $r = (s_r, e_r, t_r)$.
Each segment $j$ in $C$ is described by $c = (s_j, e_j, t_j)$.
We identify the smallest index $j$ such that $e_r \leq e_j$. We split $C$ into two parts $C_{prev} = {d_1, d_2, ... d_{j-1}}$ and $C_{after} = {d_j, d_{j+1},...., d_n}$ and insert $r$ in between, resulting in $$C_{prev} + {r} + C_{after}$$ If $e_r \geq e_n$ we append $r$ to the end of $D$.


\begin{figure}[t]
\centering
\includegraphics[width=1\columnwidth]{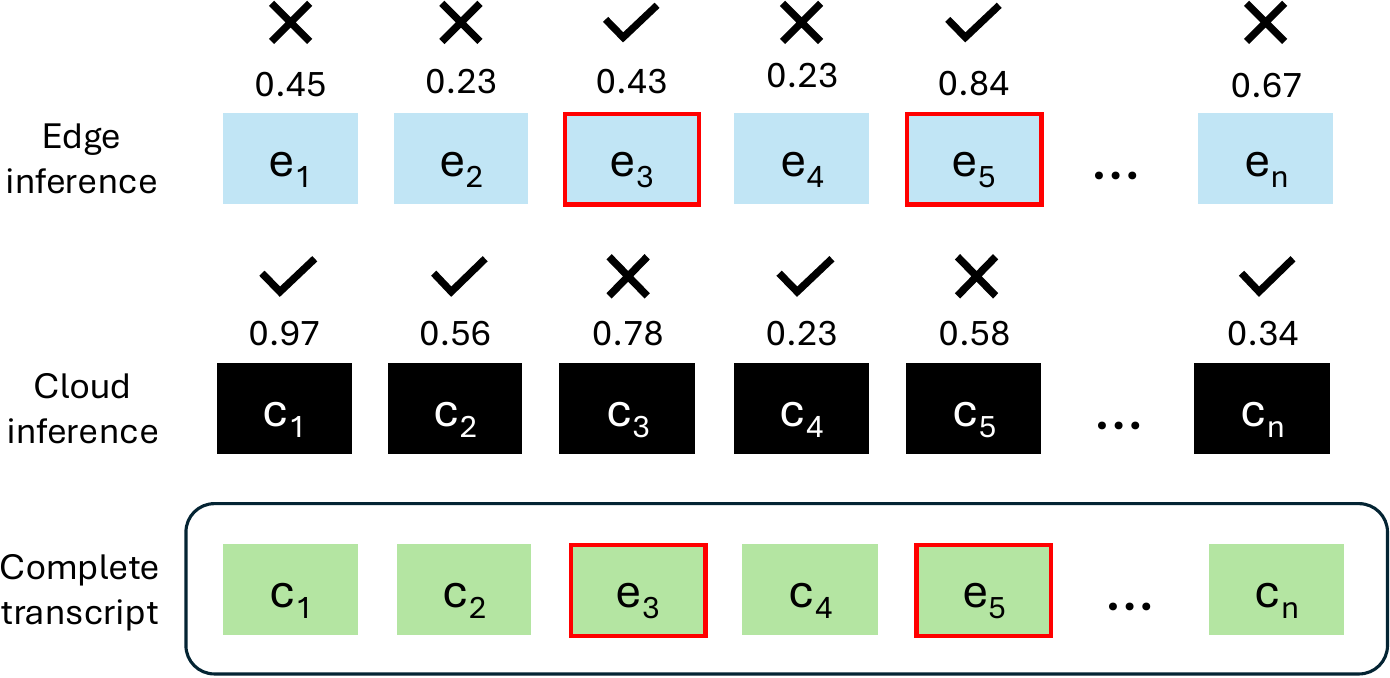}
    
    \caption{Confidence Score based transcript recovery. Red box are entity segments. Tokens in final transcript are chosen based on their confidence score given certain conditions.}

    \label{fig:cs}
\end{figure}

\subsection{\textbf{Approach B: Confidence Score Based Recovery}}
\label{sec:cs}
Timestamp based recovery might not always be accurate -- cloud predicted and edge predicted entity segments might not align because of the nature of token-level timestamp generation. A better alternative is determining which segments to keep based on the confidence score of each token. 
It is illustrated in Figure \ref{fig:cs}.

\textbf{(1) Merge \& Sort:} We merge both the cloud $C = \{ c_1, c_2, ..., c_n\}$ and edge prediction $E = \{e_1, e_2, ..., e_n\}$ and sort in a sequential manner based on their start times; $D = \{d_1, d_2, .... d_n\}; d_i \in  C \cup E$.

\textbf{(2) Overlap Detection}:
For each consecutive pair $d_i, d_{i+1}$ in $D$ we check for overlap. 
If end time of $d_i$ is greater than start time of $d_{i+1}$ i.e
$d_{i}^{end} > d_{i+1}^{start}$ there's an overlap between consecutive entries.
We can choose only one of two such overlapped consecutive entries, and we resolve by selecting the segment with the highest confidence or based on source preference.


\textbf{(3) Token Retention:} 
Among the overlapped segments, the best one needs to be chosen. Three cases are possible:
Case (a) Both segments are from cloud inference (b) Both segments are from edge inference (c) One segment is from cloud inference, the other is from edge inference.

For (a) and (b), we keep the segments with the higher confidence score. For case (c), if the segment in question is a locally saved entity segment, without any doubt, we retain the edge predicted entity segment. The rationale is that, since the entity segments were masked before offloading, whatever pseudo entity the cloud comes up with, will not be present in the ground truth transcript. 

Furthermore, we introduce a confidence threshold value $\delta$ for selecting between overlapping segments. If edge confidence score is ahead of cloud confidence score by $\delta$, we keep the edge prediction meaning - between non-overlapping segments, edge-specific data is favored if its confidence score exceeds a certain threshold, demonstrating the function’s bias towards edge-derived tokens under certain conditions.





\section{Key optimizations}
\label{sec:opt}

To enhance \sys{} inference we perform:

\textbf{Edge threshold.}
We observe that the on-device model, although poor at transcription, occasionally produces acceptable outputs.
We determine the 'acceptability' of the output transcription by looking at the confidence score of each token. For a given input, we find the average confidence score of all tokens. This average is the edge predicted threshold - the higher the threshold the better the chances of acceptability. As an optimization, if the threshold is greater than an empirically obtained value, we do not offload and choose to accept/trust the on-device inference.
There is an optimal point of this threshold value that determines the amount of acceptable locally processed inputs without compromising too much on the accuracy. \S\ref{sec:hyperparam} reports the performance of \sys{} under varying thresholds.

\textbf{Timestamp deviation.}
Per word timestamp generation is discussed in \autoref{subsec:back}.
Any misalignment of timestamps can cause the actual entity segment to not be masked, leading to a privacy breach. 
Thus while masking entity segments, we keep a buffer of 200 ms (100 before and after each segment), so that unwanted privacy leakage due to timespan misalignment can be mitigated.

\textbf{Model finetuning.}
We use a frozen pretrained Whisper-Tiny model as our on-device tiny foundation model (FM). 
We further finetune Whisper-Tiny in a parameter efficient manner using Low-Rank Adaptation (LORA) \cite{hu2021lora} on our evaluation dataset. From Table \ref{tab:et}, we can see the benefit of finetuning with an increase in accuracy by 31.89\%.

LORA freezes the pre-trained model weights and introduces small, trainable rank decomposition matrices into each layer of the Transformer architecture. This significantly reduces the number of parameters that needs to be trained for downstream tasks. In our case, the total number of trainable parameters were 589,824 out of 38,350,464 - this is only 1.58\% of the total number of parameters.
We configure LORA with a rank of 32, dropout of 0.05.
Training batch size is 16 with learning rate $1e^{-3}$ and trained for 5 epochs. We enable mixed precision training for faster computation.

\textbf{Masking configuration.}
We experiment with different choice of masks from \S\ref{sub:di} and empirically  find that while noise is the better mask configurator; cloud language models can recover well from noise-masked inputs.

\textbf{Voice activity detection and disfluencies.}
We ensure that the system actively focus on segments where actual speech occurs.
Disfluencies are interruptions in the normal flow of utterance, such as filler words ("uh," "um"), repetitions, pauses etc. Our system is configured to identify and possibly handle these disfluencies during speech recognition.

\textbf{Additionally.}
While patching, we put more faith in cloud inference, as the cloud is likely to be more right. We discard any entity token from cloud inference and always accept the locally saved edge entity tokens (with label 1). For each timespan, local inference is chosen over cloud only when the cloud confidence score is considerably low (\S\ref{sec:recovery}).

\section{Evaluation}
\label{sec:eval}
\subsection{Methodology}
\label{sub:m}
\textbf{Test platform.}
We implement \sys{} atop PyTorch 2.2.1 and Python 3.10.7.
We use an off the shelf 64 bit Raspberry Pi 4B with Quad core Cortex-A72 (ARM v8) processor having 8 GB ram as the edge device. End to end latency is calculated using the CPU execution time of this edge hardware. The rest of the evaluations are conducted on a M1 Macbook Pro. 
Model training and fine-tuning are done on an NVIDIA A40 GPU. 
We use the PyDub \cite{robert2018pydub} library for handling silence and noise in audio processing. A predefined melody segment and dummy word utterances are saved locally.

\textbf{Generating ground truth word-level timestamps.}
Using an off the shelf Whisper-large-v3 model, we generate text transcripts. We use the \textit{whisper\_timestamped} \cite{lintoai2023whispertimestamped}
library to infer per word timestamps from output cross-attention scores. An off the shelf entity extraction model \textit{en\_core\_web\_lg}
from SpaCy \cite{honnibal2020spacy} is used to derive the entity and their BIO tags from the text transcripts.  We do a reverse map in the per word timestamp table and obtain a mapping between each entity word and their corresponding timestamps (\S\ref{subsec:back}).

\textbf{Measurement.}
For the cloud runtime, we use an x86 server in lab. 
To measure the round trip time (RTT), we invoke Microsoft Azure speech service\footnote{\href{https://azure.microsoft.com/en-us/products/ai-services/ai-speech}{azure.microsoft.com/en-us/products/ai-services/ai-speech}} with the benchmark inputs. This RTT is the offload delay and is inline with prior work having an average of 900ms for a 3 second audio (stddev: 100ms), and real time factor (RTF) at 0.29–0.34 \cite{benazir2023leveraging, porcheron2018voice, potdar2021streaming}. 

\textbf{Choice of Mask.}
(1) \textit{Silence}
The duration of absolute silence must match the duration of the entity to be masked. Careful consideration is needed to ensure that silencing does not remove context that is necessary. One concern is that large language models are trained to identify silence as the end of a sentence. Transcription can abruptly stop after encountering silence.

(2) \textit{White noise} is a random signal that maintains the same intensity across all frequencies, resulting in a constant power spectral density\footnote{https://en.wikipedia.org/wiki/White\_noise}. White noise samples are typically drawn from a uniform distribution.

(3) \textit{Melody} is a combination of pitch and rhythm\footnote{https://en.wikipedia.org/wiki/Melody}.
We keep a local audio file that contains melody and cut a fixed duration of segment of it to replace the entity segment.

(4) \textit{Dummy word injection} We keep a local library of audio files containing dummy words ('Mark' etc.). The dummy segments are added to the 'to be masked' portion of input in the speech domain. The dummy segments does not have any impact on the final WER as \sys{} can exactly identify and remove all such dummy segments from the final transcript.

\textbf{Datasets.} are summarized in Table \ref{tab:ds}.

\begin{table}[th]
    \centering

    \caption{Datasets in experiments. See §6.1 for details. 
    }
    \vskip -0.10 in
    \includegraphics[width=0.85\columnwidth]{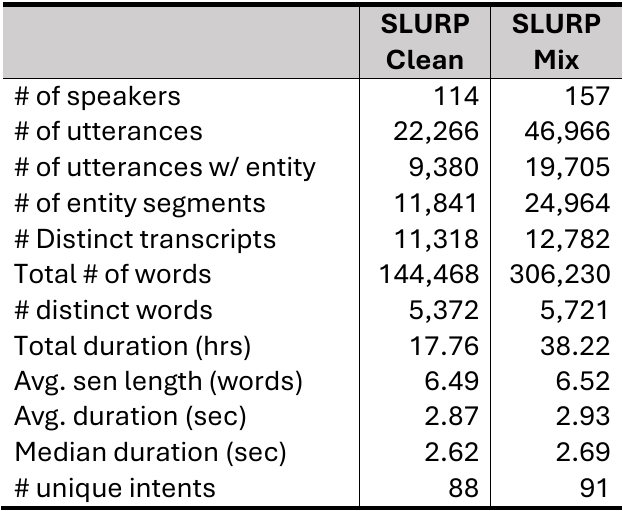}
    \label{tab:ds}
\end{table}

(1) \slurp{} is a subset of the challenging speech benchmark SLURP \cite{bastianelli2020slurp} that comprises of lexically complex and linguistically-diverse utterances such as "What is on the agenda for the four pm meeting with Joe?". The original SLURP dataset includes recordings captured in both close range (using a headset) and far range. We separate these recordings: \slurp{} consists of 22k audio samples captured with a headset including about 9k utterances containing entities.

(2) \slurpMix{} comprises of both close and far range audio recordings, amounting to a total of 47k samples from 157 speakers. It has 12k distinct transcripts and about 19k utterances containing entites.

(3) Additionally, to make a direct comparison with Preech, we experiment on VCTK-small, a subset of the VCTK \cite{veaux2017cstr} speech dataset comprising of 1249 utterances from 3 randomly selected speakers (p297, p294, 299).



\textbf{Preech.}
We compare \sys{} with Preech \cite{ahmed2020preech}, another privacy preserving speech transcription framework.
Preech does not meet the on-device requirements of our work, as its edge components are $>1.5$ GB. As it is not open source hence for our evaluation purposes we reimplement it's functionalities to the best of our abilities, ensuring that we accurately replicate its core features. 
Reimplementation details is below: 

Following the original methodology we (1) use an offline model for ASR (2) SpaCy \cite{honnibal2020spacy} to perform Named Entity Recognition (NER) on the text transcript obtained from the ASR (3) Pydub's \cite{robert2018pydub} DETECT\_NO\_SILENT()
function to segment the audio file based on silence; thus obtaining word-level timestamps. (4) Dummy words are generated using GPT-2 \cite{radford2019language} to hide sensitive entities. 
(5) A powerful cloud model for inference on obscured speech. (6) An segment reordering based approach to combine both offline and online inferences.



\begin{table}[th]
    \centering

    \caption{WER of different Whisper models.}

\includegraphics[width=\columnwidth]{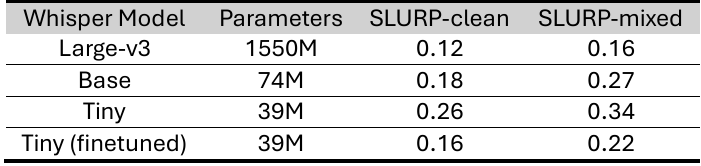}
    \label{tab:bwer}
\end{table}
\textbf{Whisper.}
Table \ref{tab:bwer} shows the word error rate (WER) performance of the whisper model family on \slurp{} (range:0.12-0.26) and \slurpMix{} (range:0.16-0.22). 
We experiment with both the zeroshot and finetuned on-device models.

Whisper-large-v3 \cite{radford2022robust} is a vanilla encoder-decoder transformer with 32 encoder-decoder layers, 20 attention heads and having a width (hidden state) of 1280. The tiny model has 39M parameters, 4 encoder-decoder layers, with 6 attention heads and a width dimension of 384.

The large-v3 model is 2.19x better than onDevice-Whisper(tiny) for \slurp{} and 2.17x better for \slurpMix{}. But for a finetuned onDevice model, it is only 1.34x and 1.40x better (for \slurp{} and \slurpMix{} respectively) validating the necessity for finetuning. We experiment with both the zeroshot and finetuned on-device models.

\subsection{Baselines}
\textit{\textbf{All-offload.}} We select Whisper-Large-v3 (1.5B) as our cloud speech service as it is the largest and most powerful open-source ASR model available to us. 
It is to be noted that, in our experiments, we consider the cloud ASR transcripts as the gold standard as the original SLURP transcripts are error-prone. 
Table \ref{tab:cs} shows the word error rate (WER) comparison of some popular cloud speech services on SLURP ranging between 0.09 to 0.46.

\textit{\textbf{OnDevice.}} are models that can be run completely on device.
(a) \textit{\textbf{OnDevice-Conformer(s)}}, a Conformer-CTC model with only 13M parameters \& 46.09 MB size is a non-autoregressive, tiny version of a Conformer model \cite{gulati2020conformer} that uses CTC loss in decoding; ASR WER is 0.4927 on \slurp{}.
(b) \textit{\textbf{\WhisperTiny{}}}: Whisper-Tiny model with 39M parameters and 75.57 MB disk size.
(c)
\textit{\textbf{\WhisperBase{}}}
Whisper-Base with 74M parameters \& 145.26 MB disk space is our largest on-device model. It is the next larger model after Whisper-Tiny in the Whisper model family.

\textit{\textbf{\preechOrg{}}} and \textit{\textbf{\preechWhisper{}}} represent two reimplementations of the Preech framework.
\preechOrg{} stays true to the original model's components and structure where Deepspeech (v0.9.3) \cite{hannun2014deep} is the on-device model and Google Speech-to-text \cite{Google_speech_to_text}
is the cloud. Meanwhile, as a more challenging alternative, \preechWhisper{} replaces the offline and online ASR models with more powerful ones such as Whisper-Tiny for on-device and Whisper-large-v3 for cloud.

\textit{\textbf{SILENCE}} \cite{cai2024lightweightprotectionprivacyoffloaded} is intended for intent classification (IC) and not ASR, nonetheless we still take it into consideration as it addresses speech content privacy.


\textit{\textbf{Ours-\{zero-shot$\mid$finetuned\}}} is our system with a zero shot Whisper-Tiny or a finetuned Whisper-Tiny model in combination with our trained NER model; noise is used as mask, Whisper-large-v3 is the cloud.


\subsection{Evaluation Metric}
\begin{figure}[t]
\centering
\includegraphics[width=0.9\columnwidth]{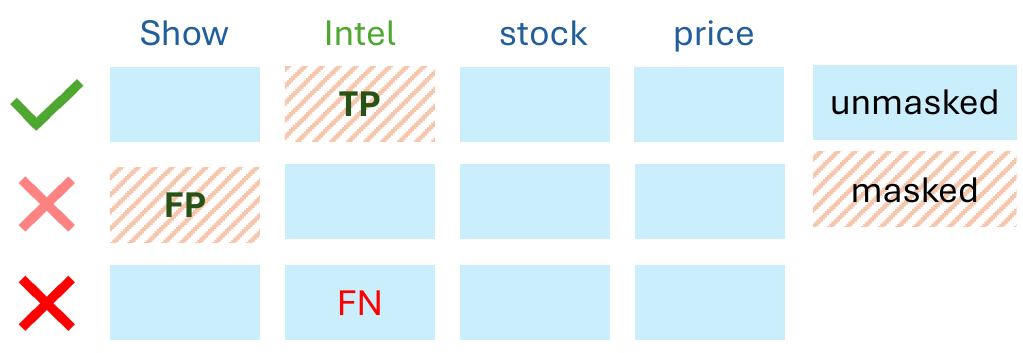}
    
    \caption{Confusion matrix of token level masking errors. For the entity 'Intel' a TP (true positive) is when that token is masked, FP (false positive) is when another non-entity token is masked and FN (false negative) is the failure to mask the entity token.}

    \label{fig:cm}
\end{figure}

Evaluating audio privacy is challenging as there is no universal standard metric. True positive (TP) is the percent of entities filtered and False negative (FN) is percent of entities that are leaked. False positive (FP) denotes excessive masking but its importance is low in our evaluation as there is no harm in masking excessively as long as the cloud model can recover from it. Illustrated in Figure \ref{fig:cm}. We aim to standardise content privacy evaluation:

(1) Token based privacy: We compare the cloud recovered transcript on masked audio with the ground truth. If original entity words are present in cloud prediction, it is a privacy leak (FN); otherwise its filtered (TP). Prior work \cite{ahmed2020preech, wang2024privacy} also follows this approach.

(2) Timestamp based privacy: For each entity timespan in utterance we concur whether that timespan (with a deviation of 200ms) is predicted as an entity or not. If it is identified as an entity, we can say with certainty that, that timespan is masked. This is a 'hard' approach as it depends on timestamp alignment of the predicted timestamps with ground truth timestamps.

\textbf{\textit{Word Error Rate (WER) for transcription correctness}} is a standard metric for ASR inference based on string edit distance ([0, 1]; lower=better) \cite{WER}. To minimize penalization of non-semantic differences Whisper Normalizer is used.

\begin{table}[t]
    \centering
    \caption{WER comparison of cloud ASR models on SLURP.
    }
\includegraphics[width=\columnwidth]{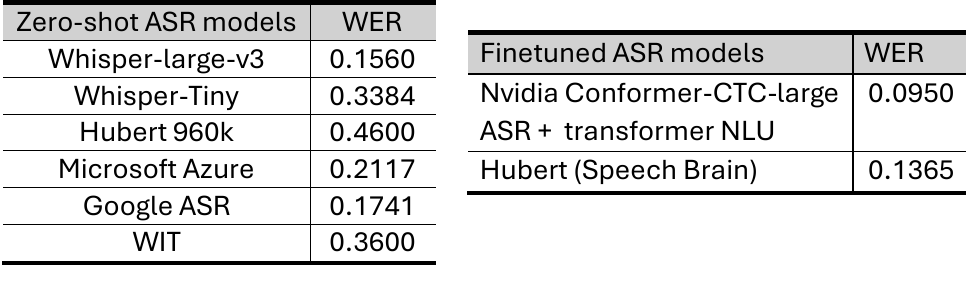}
    
    \label{tab:cs}
\end{table}
\subsection{End to end results}
\label{sec:e2e}

\begin{figure}[t]
\centering
\includegraphics[width=\columnwidth]{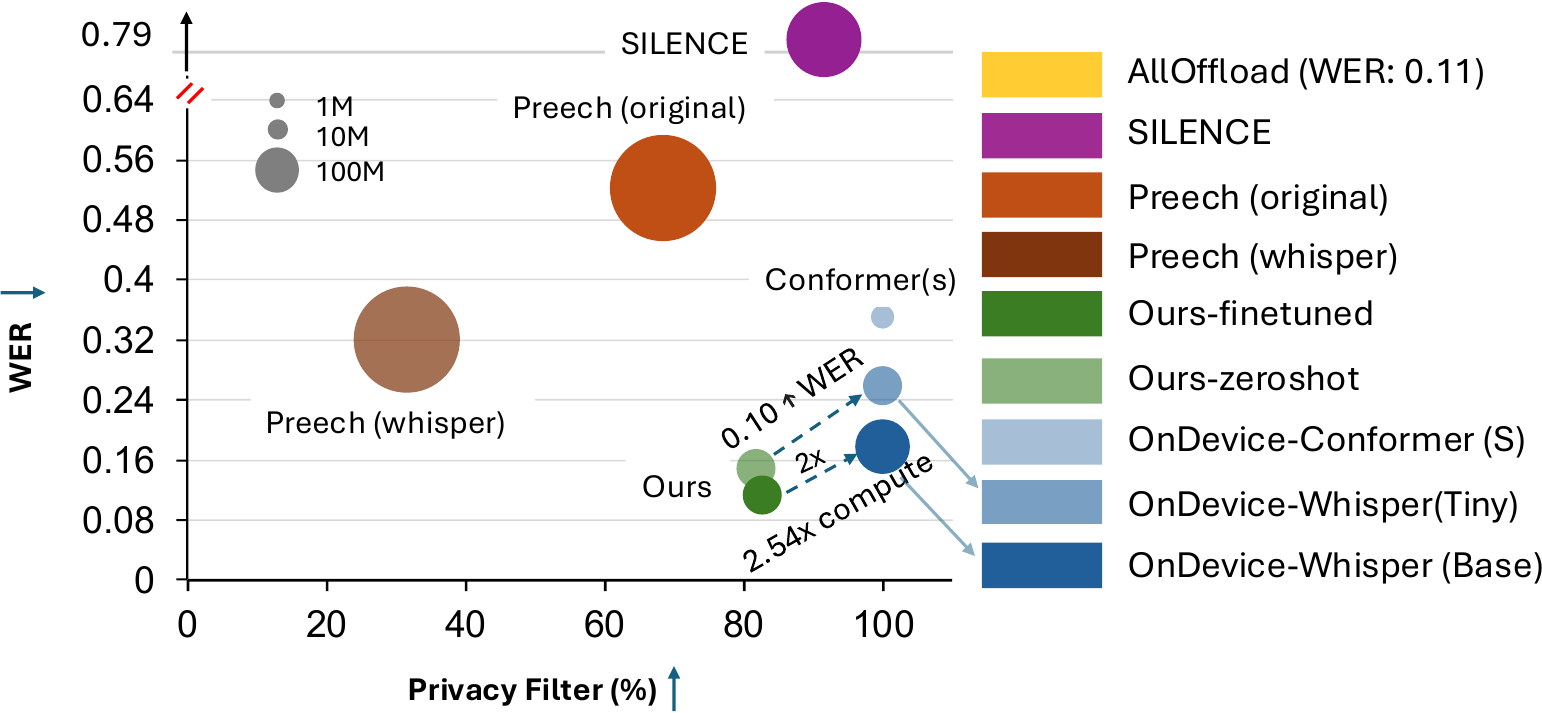}
    
    \caption{Comparison of \sys{} with baselines on \slurp{}. Whisper-large-v3 is the cloud. Lower WER indicates better performance while higher filter rate indicates stronger privacy. All \textit{OnDevice} models do not leak any entities and \textit{AllOffload} leaks 100\% of the entities. Ours-finetuned filters $\sim$83\% of entities before offloading ensuring 89\% accuracy (on-par with cloud). 
    }

    \label{fig:ce}
\end{figure}

We evaluate how well \sys{} meets the design objectives of \S\ref{sub:di}.
Unless specified otherwise, we use the default configuration of \oursfinetuned. 
As shown in Figure \ref{fig:ce}, our system offers a competitive privacy/accuracy tradeoff. 



\textbf{How well does \sys{} protect privacy?} \sys{} filters 82.69\% of the sensitive entities on-device for \slurp{} in token based privacy benchmark. For \slurpMix{} entity filtering declines by about 0.10. 
Compared to \preechWhisper{} that filters only 31\% of inputs in \slurp{}, \sys{} is 2.62x more privacy preserving. \preechOrg{} offers 14.37\% less privacy protection than ours
and also requires $> 1.5$ GB memory space - a significant amount for an on-device model. SILENCE, meant for IC, with its configurable mask generator protects 91\% of inputs. It can achieve this level of privacy as its downstream task is Intent Classification (IC) and not ASR. All OnDevice models of varying resource requirements ensure privacy at the cost of transcription accuracy. As timestamp based privacy is more rigid, our filter rate decreases by about 6-8\%. 

\textbf{Can \sys{} recover complete transcripts after masking?}
\sys{} recovers transcript with only 0.1129 WER - a significant higher accuracy compared to other baselines.
\WhisperBase{} requires 2.54x more compute than \sys{}, while \WhisperTiny{} that is of similar compute to \sys{} has 10\% higher WER than \ourszeroshot, 12\% higher WER than \oursfinetuned (\S\ref{fig:ce}). WER of \preechOrg{}, \preechWhisper{} and SILENCE are 0.52, 0.32 and 0.79 respectively. On VCTK-small, \sys{} achieves a WER of 0.05, similar to \preechOrg{}.

\textbf{How does \sys{} fundamentally differ from Preech?}
\sys{} can filter more entities as it can better predict the entity time boundaries and mask them accordingly. Unlike Preech, which relies on potentially error-prone on-device transcriptions for entity prediction, \sys{} utilizes a lightweight NER model to label entities from input tokens.
Additionally, Preech uses periods of silence to segment words, while \sys{} leverages cross-attention patterns to generate token-level timestamps and Dynamic Time Warping (DTW) for alignment to determine precise word boundaries.

\sys{} is superior to Preech in terms of transcription accuracy due to our confidence score based recovery approach.
Preech, designed for long-form utterances, assigns an order ID to each segment (about 6 seconds containing multiple words) and keeps track of it to reorder it later. This method is trivial as it lacks the precise timestamp alignment needed for accurate word-level transcription.


\begin{figure}[t]
\centering
\includegraphics[width=\columnwidth]{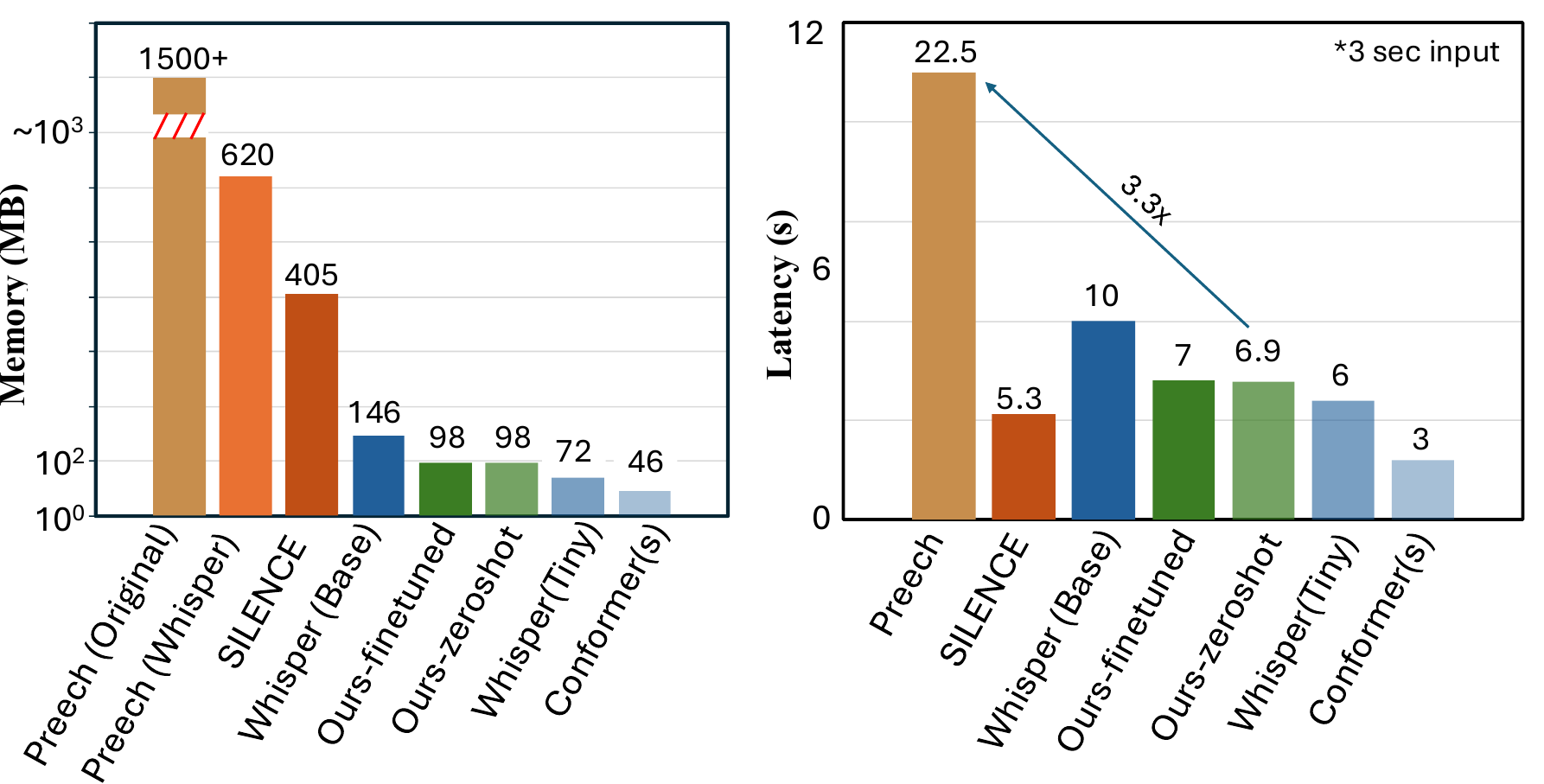}
    
    \caption{Resource cost of on-device models.}

    \label{fig:rc}
\end{figure}

\textbf{Is \sys{} compute and memory efficient?} Yes, \sys{} is 93.47\% smaller in memory (\S\ref{fig:rc}) and 17.12x less expensive than \preechOrg{} in terms of compute (\S\ref{tab:cc}). It is also 3.36x faster than the on-device \preechOrg{} model as evident from Figure \ref{fig:rc}. \WhisperBase{} is 2.54x more compute expensive. 
Further details in \S\ref{sec:o}.

\textbf{How does the NER model perform on its own?}
Our trained NER model has an accuracy of 0.9129 on \slurp{}. Accuracy is heavily penalized for a single mismatch in position of token labels. Details are reported in Table \ref{tab:ner}. 

\begin{table}[th]
    \centering

    \caption{Standalone performance (classification accuracy) of the biLSTM NER model trained and evaluated under different settings. \textbf{Takeaway: Diverse train set helps improve performance.}}
    \vskip -0.15 in
\includegraphics[width=\columnwidth]{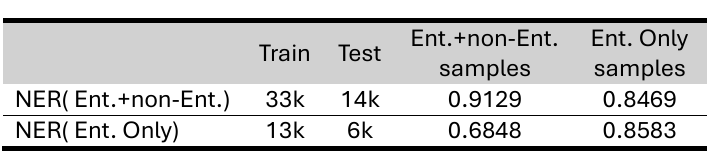}
    \label{tab:ner}
\end{table}




\subsection{Sensitivity to hyperparameters.}
\label{sec:hyperparam}

\begin{table}[t]
    \centering
    \caption{Impact of finetuning and sensitivity to edge threshold as evaluated on \slurp{}. Finetuning improves \sys{}'s WER performance by 31.89\%. Finetuning does not affect the filtered privacy. \% of local inference is inversely proportional to the edge threshold value. *Threshold=1 means all inputs are offloaded.}
    \includegraphics[width=\columnwidth]{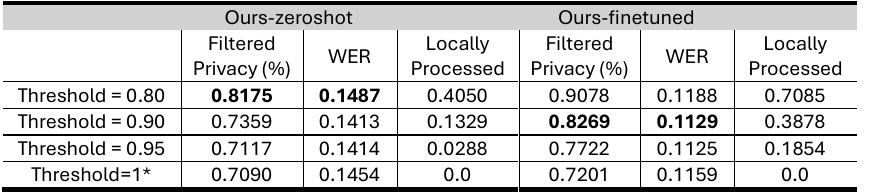}
    \label{tab:et}
\end{table}
\textbf{Impact of edge threshold.} is crucial in deciding the privacy filter rate. 
\oursfinetuned{} processes 39\% of its inputs entirely on-device, filtering 10\% more entities (from 0.72 to 0.82) when the on-device model is trusted upto a threshold value of 0.9 (ref  Table \ref{tab:et}). 
This value is chosen empirically and we can conclude that a threshold of 0.8-0.9 is optimal.

\textbf{Impact of masking configuration.}
Based on our ablation study, the best configuration is when a noise segment is used as a mask, with WER being the lowest at 0.1129. 
Our inference focuses on short-form transcripts, and masking is applied even to audio segments $< 500$ ms. In these smaller, token-level segments, inserting melody or dummy segments are less effective as they cause incomplete masking.  The different masking conditions contribute to 1-4\% to the WER.

\textbf{Impact of Finetuning}
Finetuning increases WER performance by 31.89\% compared to a non-finetuned on-device model.
Table \ref{tab:et} shows the impact of finetuning in overall performance. Finetuning does not affect privacy filtering. 

\textbf{Choice of on-device model.}
Whisper-tiny is the optimal choice for an on-device model that fulfills our requirements. 
Table \ref{tab:pe} outlines the tradeoff between performance (filtered privacy, WER) of \sys{} with different choice of on-device models. 

\begin{table}[t]
    \centering

    \caption{Tradeoff between onDevice resource consumption (w/o optimization). OnDevice Tiny offers competitive filter rate and is 30x smaller than the larger on-device models.} 
\includegraphics[width=\columnwidth]{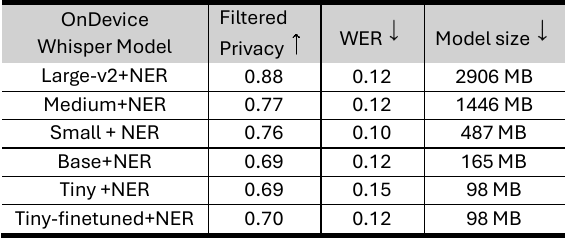}
    \label{tab:pe}
\end{table}

\textbf{Choice of Recovery Approach.}
Confidence score based transcription recovery (WER: 0.1443) is 3\% more accurate than timestamp based recovery (WER: 0.1783) for \ourszeroshot. This is expected as aligning overlapped segments and 'patching in' replacements is not straightforward.
%

\subsection{\textbf{Downstream Task: Intent Classification}}


\begin{table}[t]
    \centering
    \caption{Intent classification (IC) on \slurp{}. \sys{} can perform IC with close enough accuracy to the cloud and significantly better than the other available on-device models. *AllOffload is the gold accuracy.  
    }
\includegraphics[width=\columnwidth]{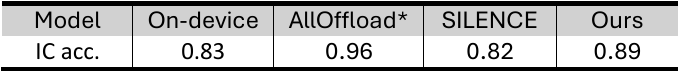}
    \label{tab:eic}
\end{table}
Aside from text transcription, \sys{} can perform intent classification (IC) with an acceptable accuracy of 0.89 from the masked speech (ref. Table \ref{tab:eic}).
The comparative on-device model that has no entity leak is a conformer(S) for ASR and mobileBert for NLU; has an accuracy of 0.83 whereas the gold cloud (Hubert) accuracy is 0.96. SILENCE is a privacy preserving approach to IC, accuracy at 0.82 protecting 91\% privacy. Ours IC accuracy is closer to the gold accuracy.

\subsection{Overhead analysis}
\label{sec:o}
\textbf{Computation Efficiency.}
To determine computational efficiency we use three metrics (1) FLOPS (2) Disk Memory Usage (MB) (3) Real-time factor
(RTF): ratio of CPU wall clock time and input duration (Ref. Figure \ref{fig:rc} and Table \ref{tab:cc}).

\begin{table}[th]
    \centering
    \caption{Computation budget for a 4s input having 18 tokens. \#params, \#FLOPS/token and RTF values are obtained from their respective test platforms.}
    
\includegraphics[width=\columnwidth]{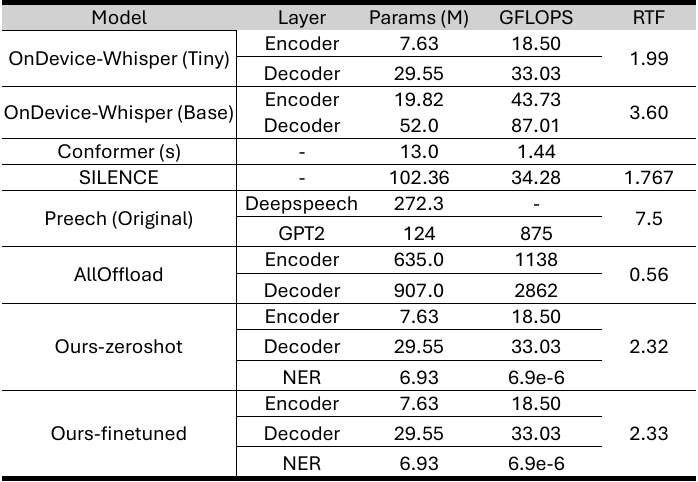}
    \label{tab:cc}
\end{table}

Ours is 93.47\% smaller than
\preechOrg{} as evident in Figure \ref{fig:rc}. The only other on-device model with competitive accuracy as ours: \WhisperBase{} has 2x the number of parameters as ours. For a 4 sec audio having 18 tokens, ours is 17.12x and 2.54x 
less compute expensive than \preechOrg{} \& \WhisperBase{}. 
\cloud{}, designed for accuracy requires computation in the range of TeraFlops; highly expensive. 
Our RTF is acceptable; latency is 2.24x and 1.55x faster compared to \preechOrg{} and \WhisperBase{} respectively. 


\textbf{Latency.}
We take into account the test server’s round trip time (RTT) in addition to the on-device inference. Given the resource constraints and processing speed of a Cortex-A72 processor, a single 3 second speech inference normally takes $\sim$6 second, our framework introduces an additional 1 second only (ref. Figure \ref{fig:rc}), we believe this added latency is acceptable given the processing limit and privacy protection.

\begin{table}[t]
\centering
\caption{Energy consumption as measured using a JouleScope JS220 Precision DC Energy Analyzer for a $\sim$6s input.}
\includegraphics[width=\columnwidth]{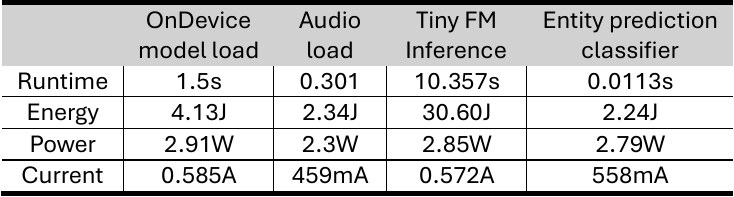}

    \label{fig:ec}
\end{table}

\textbf{Energy Consumption.}
\sys{} consumes 3 W power during inference of a $\sim$6 second speech input. The model runtime also includes the model load time and I/O operations. We measure the power and current consumption of a single inference on our test platform using a JouleScope JS220 Precision DC Energy Analyzer (ref. Figure \ref{fig:ec}). For running inference on a 6 second audio, at 5V, average power consumed is 3 W; 39.31 J energy is spent.
\section{Discussion}
\label{sec:discuss}

\paragraph{Limitations}
(1) \sys{} only preserves textual content privacy and is not meant for voice privacy that has been well-researched (detailed in Section \ref{sec:related}). Having said that, existent speech anonymization techniques can be applied if needed. (2) Although acceptable, entity prediction error from the edge model is propagated in the final transcription. 



\begin{table}[th]
\centering
\caption{Word Error Rate (WER) breakdown
}
\includegraphics[width=\columnwidth]{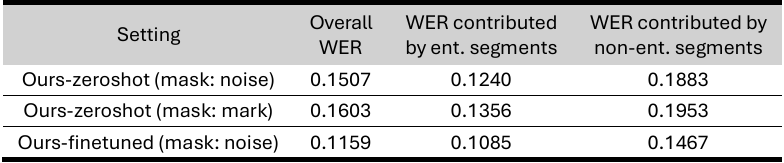}

    \label{tab:te}
\end{table}
\paragraph{Does the cloud contribute in error propagation?}
In \sys{} error contribution of edge:cloud is 1:4.
In the final transcript, WER of entity segments is 0.1085 while the non-entity segment (obtained from cloud) is 0.1467. This is understandable as the ratio of entity vs non-entity words is skewed towards non-entity words.
We can see from Table \ref{tab:te} the contribution of word segments in error propagation. 
For the best performing finetuned model, WER contributed by the entity segments (inferred by the local model) is 0.1085 whereas WER contributed by the non-entity segments (inferred by cloud and in the confidence score based recovery approach) is 0.1467.



\paragraph{Do we allow contextual inference from masked speech?} Yes, contextual information or action items (such as 'meeting' in Figure \ref{fig:o}) surrounding the masked segment is necessary in downstream tasks for deducing speaker intent or flow of action. \sys{} enables partial redaction while preserving meaning and helps in overall comprehension.
\paragraph{Cloud behavior on masked segments}
The SOTA cloud has a robust language model (LM) that can recover from the injected noise in masked speech. In places of noise, it outputs dummy entities as the LM can predict next token from context. We quantify the coherency of the cloud by comparing cloud-predicted non-entity segments with the ground-truth counterparts, observing that cloud is 78.15\% closer to ground truth on masked inputs.

\paragraph{How well does the local model perform on entity words?}
Whisper-Tiny is only 0.05 lower in WER on entity words within a transcript than \WhisperBase{}. In a complete transcript, non-entity words are significantly more than the entity words, hence WER propagated in final transcription because of on-device model ASR is miniscule.

\paragraph{Can a smaller NER model be used?} Our classifier adds minimal overhead compared to alternatives. 
Typical on-device NER models (e.g., DistilBERT-NER~\cite{sanh2019distilbert}, TinyBERT~\cite{jiao2020tinybert}, ALBERT-Base~\cite{lan2020albert}) are 2–3$\times$ larger, whereas our proposed NER model achieves 91\% accuracy with far greater efficiency for on-device inference.

\paragraph{Does \sys{} hallucinate?} \ourszeroshot{} hallucinates 0.0594 number of times as evaluated on 2760 samples. Finetuning reduces this hallucination to 0.0362, similar to \cloud{} that hallucinates 0.0322 of the time.

\paragraph{Alternate designs rejected by us}
We explore pruning the on-device tiny FM further to limit memory usage. From our experiments we find that layer pruning causes 50\% hallucinations; quantization also doesn't uphold performance; hence we discard these approaches. 

\paragraph{Possible extensions} Mitigation strategies to reduce error propagation such as temperature scaling \cite{guo2017calibration}, or label smoothing \cite{szegedy2016rethinking} are complementary and could be integrated into our framework without altering its core design. 
These serve as valuable extensions, but the lack of calibration does not diminish our main contributions.  Additionally, ensemble edge models were excluded to avoid significant memory and compute overhead.


\section{Related work}
\label{sec:related-work}
\label{sec:related}


\textbf{Speech privacy} is a relatively unexplored domain \cite{latif2020deep}. 
Most work \cite{dias2018exploring, aloufi2019emotionless, granqvist2020improving} focus on paralinguistic attributes or \textit{voiceprint} privacy in speech such as gender, age etc. but hardly tackles speech content privacy.
Prior work that addresses speech content privacy can be too limited in scope (can mask only predefined keywords) \cite{qian2017voicemask}, computationally infeasible on-device \cite{ahmed2020preech} or suited for selective downstream tasks such as intent classification (IC) \cite{cai2024lightweightprotectionprivacyoffloaded}.\cite{cheng2022personal} provides an overview of the latest developments in security and privacy for personal voice assistants. 

SILENCE \cite{cai2024lightweight} performs IC by obscuring short-term details from an input speech without significantly damaging the long-term dependency. Preech \cite{ahmed2020preech} is a system closest to ours that protects speech privacy through selective word scrubbing and GPT based voice conversion.
However, it is unsuitable for mobile deployment due to latency and the high cost of invoking multiple cloud services. \sys{} captures the problemscope of \cite{benazir2024maximizing} but is more concrete and empirical in its findings.

On the edge computing side, SODA \cite{atrey2023soda} focuses on mitigating the leakage of proprietary model information rather than user data in edge machine learning deployments, demonstrating its framework on an RPi3-based setup comparable in scale to ours.
Although both aim to preserve privacy in edge–cloud inference, \cite{chi2018privacy} prevents representation-level leakage through secure computation partitioning, whereas 
\sys{} prevents content-level leakage by masking sensitive speech entities before offloading.

\cite{olade2019smart} introduces an intermediary blocker between the user and smart speaker. VoiceGuard \cite{brasser2018voiceguard} introuduces hardware-assisted security by performing ASR in the trusted execution environment (Intel’s SGX enclave) but it is memory intensive.
\cite{sun2020alexa} introduces MicShield, a novel selective jamming mechanism - that selectively blocks private speech.
VoiceMask \cite{qian2017voicemask} implemented on Android with two sanitization phases, conceals voiceprints through robust voice conversion, performs anonymization and protects input content privacy through a keyword substitution heuristic. A distinct difference between VoiceMask and ours is that VoiceMask can only sanitize predefined keywords, whereas we do not have any limit on such keywords. 

\textbf{Voice Privacy}  Voiceprint is the distinctive identifiable pattern of each individual \cite{stoidis2021protecting}. Users are at risk of spoof attacks on speaker authentication systems, mitigating strategies typically involve speaker anonymization through voice conversion \cite{qian2018hidebehind}. But simple voice conversion is unable to effectively protect against an attacker \cite{srivastava2020evaluating}.

Approaches for paralinguistic privacy include differential privacy \cite{granqvist2020improving}, cryotographic approaches \cite{gilad2016cryptonets, riazi2019xonn, nautsch2019preserving}, secure multi-party computation (SMC) \cite{pathak2013privacy, nautsch2019preserving}, federated learning \cite{leroy2019federated} etc. These can be computationally heavy and slow \cite{nautsch2019preserving, qian2017voicemask}.
In recent days, disentangle based representation learning \cite{aloufi2020privacy, wang2024privacy, aloufi2023paralinguistic} is emerging and has been used for privacy preserving speech emotion recognition \cite{dias2018exploring, aloufi2019emotionless}, speaker verification \cite{granqvist2020improving, gong2017crafting} but they demand considerable computational resources (compute requirement is in GFLOPs and in GB memory).
Mitigating strategies typically involve speaker anonymization through voice conversion \cite{qian2018hidebehind}.
Threat model include tasks such as privacy preserving speech emotion recognition \cite{dias2018exploring, aloufi2019emotionless}, speaker verification \cite{granqvist2020improving, gong2017crafting}, gender identity \cite{stoidis2021protecting, jaiswal2020privacy, wu2021understanding}.


Traditional methods include ultrasonic microphone jammers (UMJ) to preventing illegal eavesdropping \cite{chen2022big, gao2023cancelling} but they corrupt speech semantics.
Ultrasonic microphone jammers prevent eavesdropping by exploiting the non-linearity of microphones. 
\cite{gao2023cancelling} introduces MicFrozen that cancels speech signals and adds coherent noises simultaneously.

More effort has been dedicated on protecting paralinguistic attributes than textual content \cite{gong2017crafting, abdullah2021hear, nelus2019privacy}.







Edgy \cite{aloufi2023paralinguistic}, a 134MB model performs on-device paralinguistic information filtering based on disentangled representation learning. With different cloud based services, edgy's WER is between 11.31 and 85. This is on speech filtered for paralinguistic privacy, it does not protect content privacy.
\cite{aloufi2019emotion} builds an edge-based system to filter affect patterns from a user’s voice.

\section{Conclusion}
We present \sys{}, a hybrid ASR inference engine that can filter $\sim$ 83\% of sensitive entities on-device, while maintaining state-of-the-art transcription accuracy. We extend the capabilities of tiny speech foundation models and maximize inference performance to the best of our capabilities.

\section{ACKNOWLEDGMENTS}
The authors were supported in part by NSF awards \#2128725, \#1919197, \#2106893 and \#2426353. The authors thank the anonymous reviewers and sheperd An Wang for their insightful feedback.

\bibliographystyle{ACM-Reference-Format}
\bibliography{main}

\end{document}